# A Testbed for Assessment of Fountain Codes for Wireless Channels

**M Usman, J Dunlop**
**University of Strathclyde, Glasgow**
**{mohammed.usman, j.dunlop}@eee.strath.ac.uk**

*Abstract* - **Luby Transform (LT) codes are a class of fountain codes that have proved to perform very efficiently over the erasure channel. These codes are rateless in the sense that an infinite stream of encoded symbols can be generated on the fly. Furthermore, every encoded symbol is information additive and can contribute in the decoding process. An important application of fountain codes which is being considered is the delivery of content over mobile wireless channels. Fountain codes have low computational complexity and fast encoding and decoding algorithms which makes them attractive for real time applications such as streaming video over wireless channels. LT codes are known to perform close to capacity on the binary erasure channel and it is envisaged that they would have good performance on other channels such as mobile communication channels and satellite links. This paper considers the development of a test-bed to study the performance of fountain codes over such channels. The performance of LT codes on the binary symmetric channel and Additive White Gaussian Noise (AWGN) Channel is presented as examples of the testbed usage.**

## I. Introduction

Transmission of data reliably over a communication channel has been a topic of much research. As data passes through the communication channel it is distorted by the effects of the channel and reliability is achieved by either retransmitting the erroneous data or using appropriate coding schemes. Retransmission protocols require a reverse channel and perform poorly in the case of multicast and broadcast transmissions. In such circumstances, the retransmissions would be redundant to most of the receivers and the transmitter could also suffer from what is called as 'feedback implosion'. Retransmission can also cause long delays in high load situations. Conventional error coding schemes require that the transmitter and receiver know the channel statistics a-priori in order to fix the code rate. Typically, the sender or the receiver may scan the channel and make a reasonable guess of the present statistics of the channel. However, in the case of time-varying channels such as radio channels it is necessary to make frequent updates on the channel statistics.

Fountain codes overcome both these problems – there is no need for retransmissions and there is no need for the sender and the receiver to know the channel statistics a-priori. For a given set of '$k$' message symbols, a fountain code can generate a potentially infinite stream of encoded symbols.

All encoded symbols are generated independently from each other and are therefore information additive i.e. it does not matter which encoded symbols are received as long as sufficient number of them are received. The principle of fountain codes can be thought of as analogous to solving a set of simultaneous linear equations. If there are 3 equations in 3 unknowns, the equations can be solved for the unknowns. If there are more than 3 equations available in the same 3 unknowns, any 3 equations may be chosen to solve for the unknowns [1].

A reliable decoder for fountain codes is one which can recover the $k$ message symbols using any $k' = k(1 + \varepsilon)$ encoded symbols. The factor $1+ \varepsilon$ is called the decoding inefficiency, which is the fraction of excess symbols more than $k$ that are required for decoding the $k$ message symbols. For good fountain codes $k'$ is close to $k$ i.e. $\varepsilon$ close to zero. It should be observed that for a given degree distribution, the value $k'$ is the same regardless of the statistics of the channel. Poorer channels would simply result in longer waiting times for the receiver to receive $k'$ encoded symbols. This paper presents the architecture of a testbed designed to study fountain code performance on radio channels. The binary symmetric channel and the AWGN channel are used as examples to demonstrate the testbed operation. Real channel characteristics may be incorporated by use of data derived from a soft decision Viterbi decoder [2].

## II. Fountain encoding and decoding

The degree distribution forms the key to the design of good fountain codes. The degree of an encoded symbol is defined as the number of message symbols covered by it. The degree '*d*' of each encoded symbol is chosen independently from an appropriate degree distribution. In this paper, the Robust Soliton (RSol) distribution [3] and the distribution defined by equation (1), optimised by Shokrollahi [4] are used. It must be noted however, that it is not the aim of this paper to design new degree distributions for fountain codes.

The Robust Soliton distribution $\mu(d)$ is a modification of the Ideal Soliton distribution $\rho(d)$ and is defined as follows:
Let

$$R = c.\log_e(k/\delta)\sqrt{k}$$

for some constant $c > 0$. '$c$' can be thought of as a free parameter, with a value smaller than 1 giving good results. '$\delta$' is a bound on the probability that the decoding fails to run to completion after a certain number $(k')$ of encoded symbols have been received.
'$R$' is the expected number of degree one encoded symbols throughout the decoding process.

$$\rho(d) = \begin{cases} 1/k & for\ d = 1 \\ 1/d(d-1) & for\ d = 2,3.....k \end{cases}$$

$$\tau(d) = \begin{cases} R/d.k & for\ d = 1,2.........(k/R)-1 \\ R.\log_e(R/\delta)/k & for\ d = k/R \\ 0 & for\ d = (k/R)+1..........k \end{cases}$$

$$\beta = \sum_{d=1}^{k} \rho(d) + \tau(d)$$

$$\mu(d) = (\rho(d) + \tau(d))/\beta \quad \text{for } d = 1\ldots\ldots\ldots k$$

The expected number of degree one encoded symbols for the Ideal Soliton distribution is 1 which accounts for its failure in practice. The RSol distribution ensures that the expected number of degree one encoded symbols is '$R$'.

$\mu(d)$ is the probability distribution of the degree '*d*'. $\mu(d)$ is designed such that occasional encoded symbols have high degree to ensure that there are no message symbols that are not covered during the encoding. There should also be sufficiently many encoded symbols of low degree so that the decoding can get started and keep going [5]. This is to ensure that the decoding is successful with the use of as few encoding symbols as possible.

The computation cost of encoding and decoding a LT code generated using the RSol distribution is of the order of $k.log_e(k)$ symbol operations, where '$k$' is the number of message symbols and $log_e(k)$ is the average degree of each encoded symbol. Thus the average degree of each encoded symbol and hence the computation cost grows as the number of message symbols ($k$) increases [5]. The distribution of equation (1) overcomes this difficulty by having an average degree of each encoded symbol that is constant (average degree = 5.87) and smaller than that of the RSol distribution.

$$\Omega(X) = 0.007969X + 0.493570X^2 + 0.166220X^3 + 0.072646X^4 + 0.082558X^5 + 0.056058X^8 + 0.037229X^9 + 0.055590X^{19} + 0.025023X^{65} + 0.003135X^{66} \quad \ldots\ldots\ldots\ldots\ldots\ldots\ldots\ldots(1)$$

where,

$0.007969X$ means that the probability that degree = 1 is 0.007969,
$0.493570X^2$ means the probability that the degree = 2 is 0.493570 and so on.

This simplification however leaves some of the message symbols uncovered by any of the encoded symbols and hence these message symbols cannot be recovered. An inner code could then be used to decode the uncovered message symbols.

The encoding is performed as follows:
A degree '*d*' is chosen at random from the degree distribution. '*d*' distinct message symbols are chosen at random from among the '*k*' message symbols. These '*d*' message symbols participate in the computation of the encoded symbol and are called its neighbours. The value of the encoded symbol is the XOR of these '*d*' message symbols. The encoded symbols are then transmitted over the channel. The encoding operation induces a graph connecting the message symbols with the encoded symbols. The resulting code can be thought of as an irregular low density generator matrix code.

The decoder needs to know the degree and the set of neighbours of each encoding symbol in order to decode the message symbols. In actual systems, this information can be conveyed to the decoder in one of many different ways depending on the application. The degree and list of neighbours may be explicitly transmitted along with the encoded symbol.

A key can be associated with each encoded symbol and this key may be passed to the decoder along with the encoded symbol. The decoder could then locally generate the degree and list of neighbours for the encoded symbols using the key.

In this paper, an assumption is made that the information needed to reconstruct the graph (i.e. the degree and the set of neighbours for each encoded symbol) is conveyed to the receiver by some suitable means. The decoding of LT code is performed using the Belief Propagation decoding algorithm which can be described in simple terms as follows:

If there is an encoded symbol with degree = 1, then the message symbol which is neighbour to this encoded symbol can be recovered since the encoded symbol is a copy of the message symbol.

The value of this decoded message symbol is XOR'ed with all the other encoded symbols that have it as neighbour.

The decoded message symbol is removed from the list of neighbours for each of these encoded symbols and the degrees of these encoded symbols are reduced by one to reflect this removal. For example, if there is an encoded symbol with degree = 2 that has the current decoded message symbol as one of its two neighbours, then the decoded message symbol is XOR'ed with this encoded symbol. The decoded message symbol is then removed from the list of its neighbours and its degree is decremented by one. Thus the new degree of the encoded symbol is one and has only one member in the list of its neighbours. The process is repeated until all encoded symbols of degree = 1 are exhausted. A decoding failure is reported if all the message symbols have not been recovered at the end of the decoding process.

## III Testbed

To emulate fountain codes, the experimental testbed consists of the following sections as shown in Fig 1.

### Input Buffer and Message symbol buffer

The message to be encoded is stored in the message symbol buffer. If the data to be encoded comes from a streaming source, the first set of '*k*' incoming symbols is stored in the message symbol buffer. The following symbols of the incoming stream are buffered in the input buffer while the first set of '*k*' message symbols is being processed. Once processed, the next set of '*k*' symbols from the input buffer is loaded into the message symbol buffer for processing. The size of the input buffer must be large enough to accommodate the incoming stream of symbols while the current set of '*k*' input symbols in the message symbol buffer is being processed.

**Encoder**

The encoder comprises of the degree selector, the associator and the calculator. Every time an encoded symbol is to be generated, the degree selector randomly selects a degree '*d*' for the encoded symbol according to a degree distribution. In the case of using the degree distribution of equation (1), the degree is selected by generating random numbers between 0 and 1. For example, if the random number generated lies between 0 and 0.007969, the degree is chosen to be 1 and so on.

The associator selects '*d*' distinct message symbols from the '*k*' message symbols. The associator is implemented in such a way that each of the '*k*' message symbols has a roughly even chance of being selected as a neighbour of the encoded symbol.

The calculator computes the value of the encoded symbol as the XOR of the '*d*' message symbols selected by the associator.

**Channel**

This is the communication channel over which the encoded symbols are transmitted. In this paper the binary symmetric channel and the AWGN channels are considered. Future work will consider mobile wireless channels.

**Decoder**

The decoder comprises of a buffer, a reducer and a calculator. The buffer stores the received encoded symbols along with the information necessary for the decoding i.e. the degree and list of neighbours for each received encoded symbol. Once a sufficient number of encoded symbols have been collected in the buffer, decoding is performed as described in the previous section.

The calculator computes the XOR of any recovered message symbol with the encoded symbols which have it as neighbour.

The reducer reduces the degree of the encoded symbol which is XOR'ed with any recovered message symbol by decrementing the degree of such an encoded symbol by 1. It also removes the recovered message symbol from the list of neighbours of the encoded symbol. The testbed can be used to study fountain codes over different channels using an appropriate model of the channel.

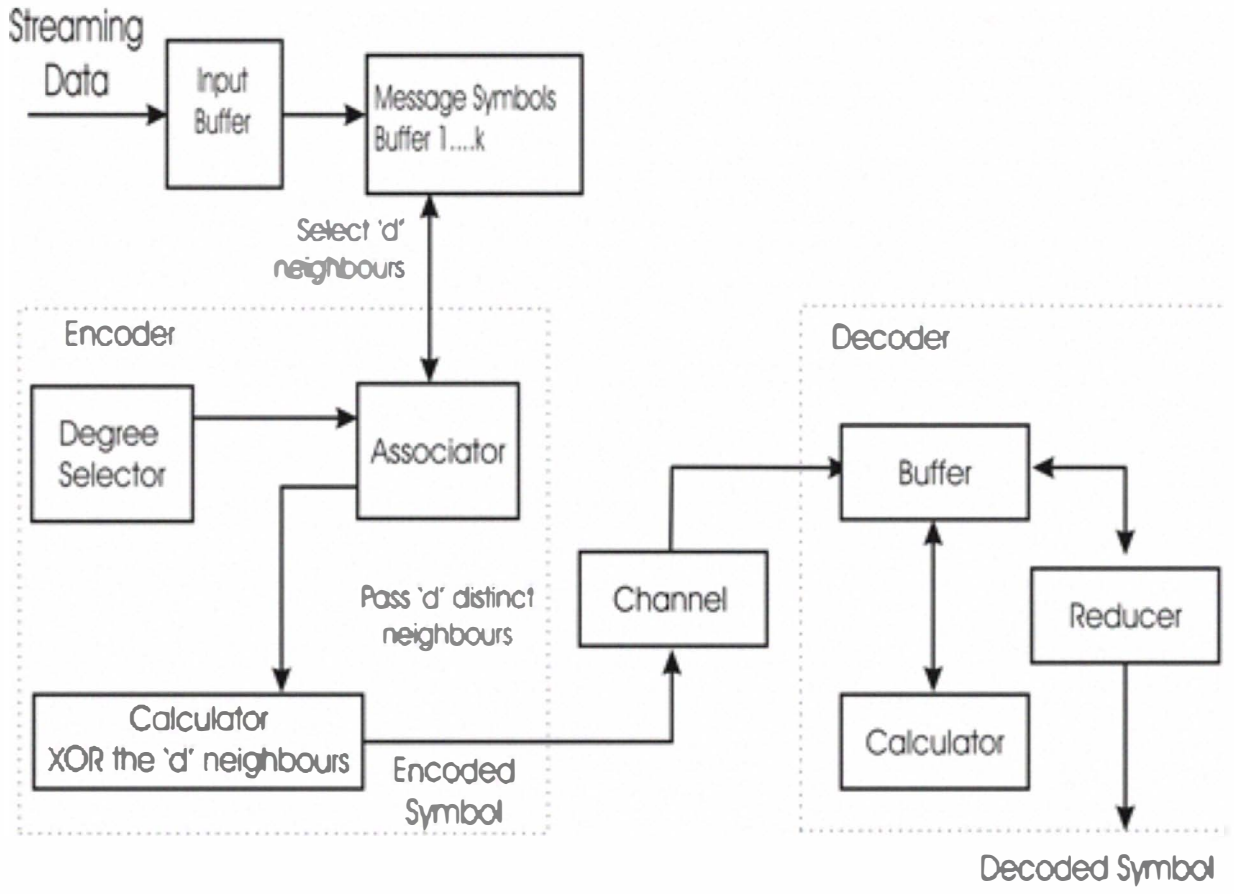

Fig.1 Architecture of the testbed

## IV LT codes performance over a perfect channel

A sequence of $k$ = 1021 message symbols is generated. The encoding process is performed as described in section II using the Robust Soliton distribution. The encoded symbols are then transmitted over the communication channel and decoding is attempted once a certain minimum number ($n = k = 1021$) of encoded symbols are received. It is a property of LT codes that very little decoding is possible until slightly more than '*k*' symbols are received. This property is clearly observed in the plot of Fig 2. The message symbols are not transmitted over the channel. In the first case the channel is assumed to be a perfect channel and introduces no errors. This idealistic assumption allows the number of additional encoded symbols that would be required to ensure that the decoding completes successfully to be observed and comparison of the results with values derived from the analysis of the Robust Soliton distribution.

If the decoding fails with $n = k$ encoded symbols, the receiver takes in more encoded symbols and attempts decoding again. This process is repeated until the decoding runs successfully to completion and all the '$k$' message symbols are recovered. The plot of Fig.2 shows the fraction of un-recovered message symbols against the number of encoded symbols used in the decoding process.

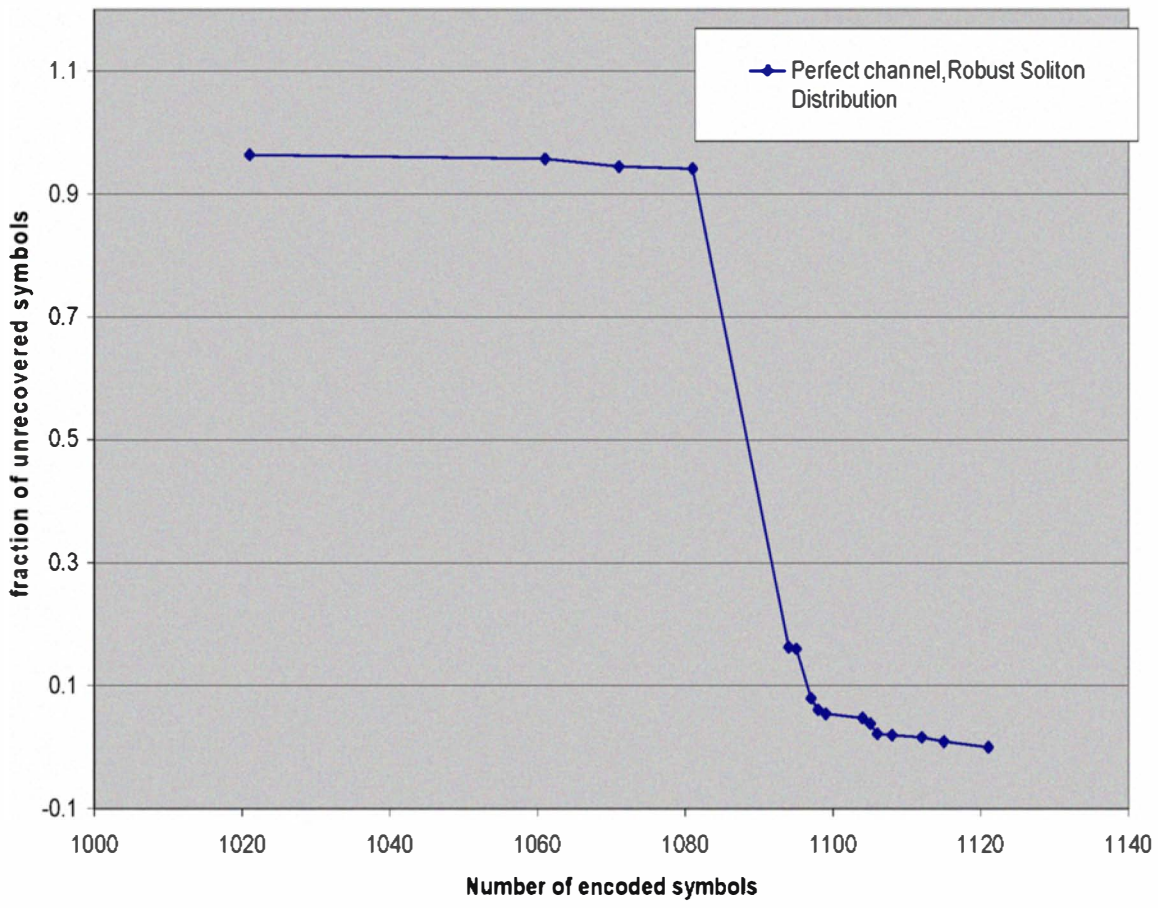

Fig.2 LT code-perfect channel constructed using the Robust Soliton distribution using c=0.01, $\delta$ = 0.5, $k$ = 1021.

Luby's analysis [3], [5], [6] shows that receiving about $k' = k + 2R\log_e(R/\delta)$ encoded symbols ensures that all symbols can be recovered with probability of at least 1- $\delta$ for an appropriate value of '$c$'. With $c = 0.1$ and $\delta = 0.5$, the decoder should recover the $k$ = 1021 message symbols from $k' \approx 1210$ encoded symbols *on average* with probability at least 0.5. It means that receiving $k'$ encoded symbols ensure the recovery of all the $k$ message symbols with probability at least 1- $\delta$.

The LT code simulation for the same settings recovers all the message symbols from $k'$ = 1198 encoded symbols *on average* which gives an overhead of about 17%. Choosing an appropriate value for '$c$', LT codes can be tuned to have overheads between 5%-10%. Setting '$c$' to 0.01, all the 1021 message symbols are recovered from 1121 encoded symbols *on average* which gives an overhead of 9.79% and with $c$ = 0.03 an overhead of 11% is incurred (i.e. $k'$ = 1134 encoded symbols *on average* are needed to recover all the 1021 message symbols).

## V. LT codes over Binary Symmetric Channel

The binary symmetric channel is simple to model and the performance of the LT code was studied on this channel for both the RSol distribution and the distribution of equation (1). The encoding is performed as described in section II with the degrees drawn from the RSol distribution. The encoded symbols are passed through the binary symmetric channel with $p$ = 0.01, where '$p$' is the bit flip probability.

Decoding is attempted after receiving a minimum of $n$ = 1021 encoded symbols. If all the message symbols are not recovered, the receiver takes in more and more encoded symbols and runs the decoding until either all the '$k$' message symbols are recovered or a certain maximum number of encoded symbols have been received. LT codes, being rateless, do not lay any restrictions on the maximum number of encoded symbols that can be generated. However, in practical systems the transmission cannot continue forever. Hence, a certain maximum number, sufficiently larger than the number of encoded symbols ($k'$) necessary for successful decoding is chosen after which decoding is terminated. The experiment is repeated by drawing the encoded symbols from the distribution given by equation (1).

Fig 3 shows the plot of the fraction of unrecovered symbols against the number of encoded symbols used in the decoding, for the codes generated using the RSol distribution and for the codes drawn using the distribution of equation (1). For comparison, the plot for the case of the perfect channel is also included on the same graph. The chain reaction property of fountain codes can be observed from the plot whereby a small set of new encoded symbols can trigger the recovery of a large number of the message symbols. All the message symbols are recovered successfully for the code drawn from the RSol distribution. However, for the code generated using the distribution of equation (1), a small fraction of symbols remain un-recovered even after the maximum number of encoded symbols has been received.

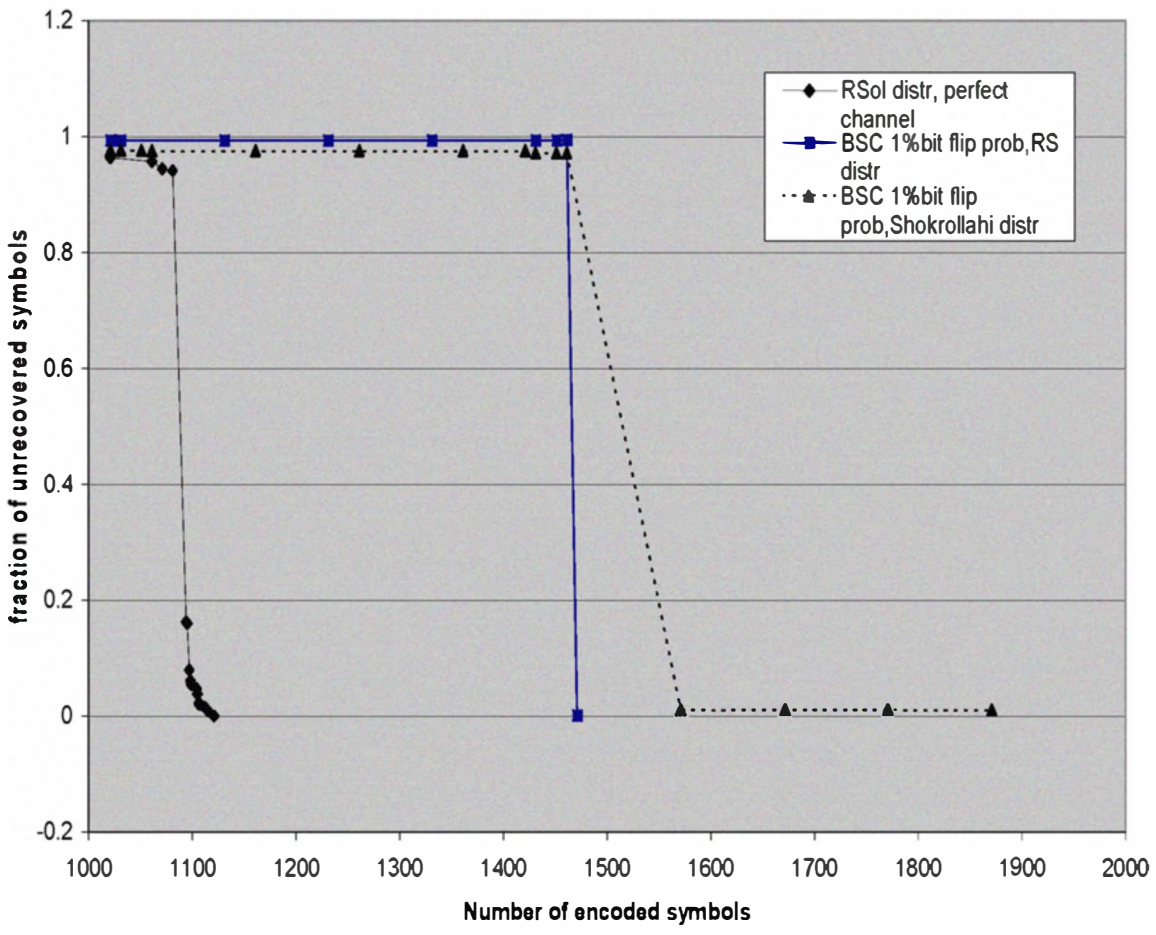

Fig.3 LT codes over a Binary Symmetric Channel

These un-recovered symbols may be attributed to the fact [7] that some of the message symbols are not covered by any of the encoded symbols or covered by only a small number of encoded symbols. It should be mentioned that the distribution of equation (1) is not designed to be used for LT codes as such, but is optimized for Raptor codes, which are LT codes with an inner code. Typically the inner code is a high rate Low Density Parity Check (LDPC) code. LT codes require highly intricate degree distributions in order to ensure that all the message nodes are covered with high probability. Raptor codes on the other hand ease the condition that all message symbols must be covered and use relatively simple degree distributions. In order to transfer large files, a Raptor code is a better choice compared to a LT code, for reasons concerning computation cost described in section II.

## VI. LT codes over AWGN channel

Fig 4 shows the performance of LT codes obtained from the testbed when the channel is an AWGN channel. In this case the codes are generated using the RSol distribution and the distribution of equation (1) over an AWGN channel with symbol energy to noise ratio equal to -2.83 dB which corresponds to capacity 0.5. The Shannon limit for this channel is 2 i.e. to decode '$k$' message symbols '$2k$' encoded symbols should suffice, ideally using random codes. The results would be indicative of how close to the Shannon limit that LT codes perform. It can be observed from the plot of Fig 4 that all '$k$' message symbols are decoded using only slightly greater than $2k$ encoded symbols which is close to the Shannon limit.

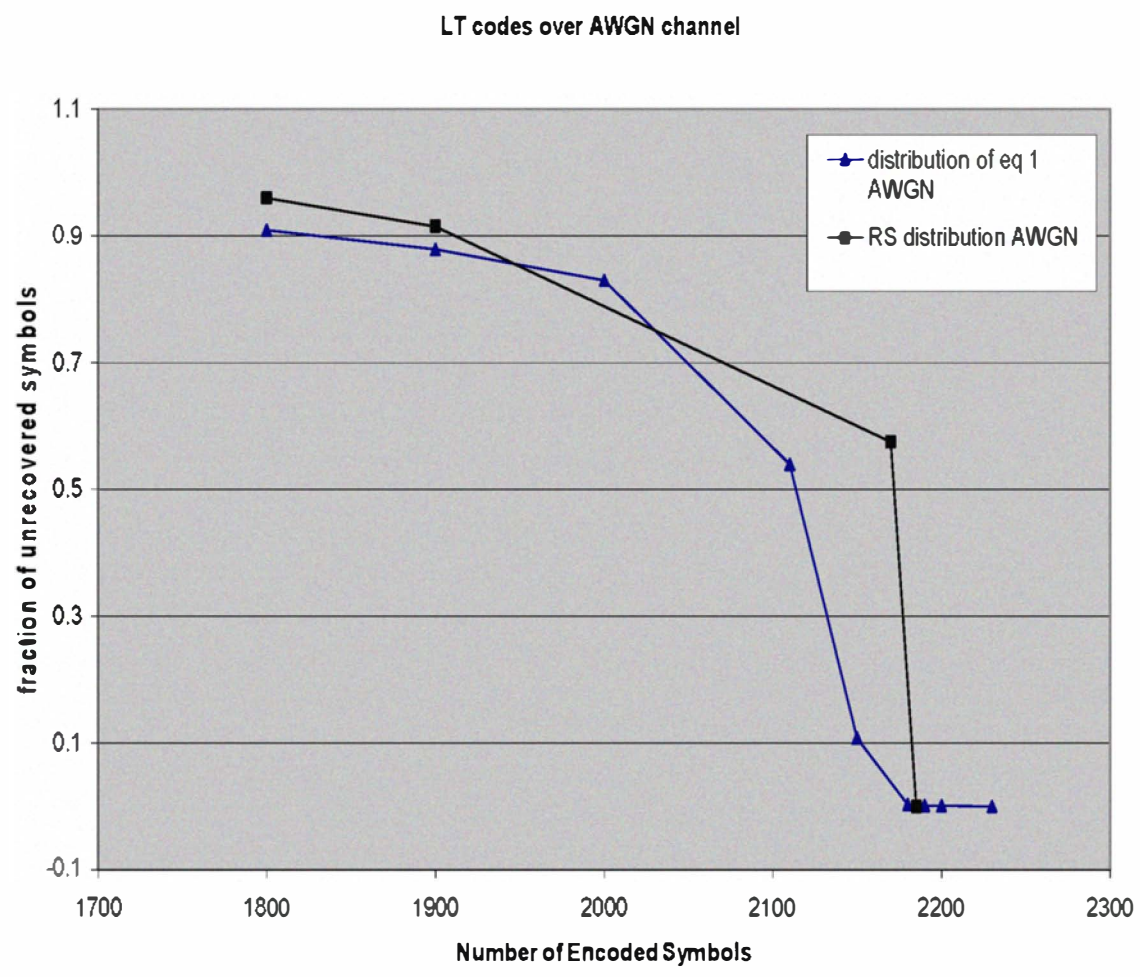

Fig. 4 LT codes over AWGN channel

The results demonstrate the progression of the decoding process as new encoded symbols are received. Comparing the plots for the RSol distribution and the distribution of equation (1) in Fig 3 and Fig 4, it is observed that the decoding requires more encoded symbols in the case of the LT code derived from equation (1).

At large values of '$k$', there could be a small number of unrecovered message symbols even with large overheads as it is probable that there are some message symbols that are not covered during the encoding process.

Fading, reflections, or other non-gaussian impairments are either dynamic or not known before the channel is engaged. Often the codes are still evaluated relative to their performance in a Gaussian channel, since whiteners and other techniques are usually used to make the non-gaussian channel look Gaussian to the coding.

Further work is aimed at the study of LT codes and Raptor codes performance over mobile wireless channels. It is aimed to incorporate an inner irregular LDPC code with left-regular (i.e. each message node having degree = 4), right Poisson distribution (i.e. check node degrees having Poisson distribution) as suggested in [4], [8]. Raptor codes are being commercially used for distribution of data over the internet but the literature on the performance of fountain codes over wireless links seems sparse.

## VII Conclusion

The overall objective is to choose fountain codes for use over heterogeneous radio channels. Of particular interest is the associated overhead as the mobile radio channel will be limited in bandwidth. A testbed has been developed to study the performance of fountain codes on various channels. The initial results over the binary symmetric channel and the AWGN channel show that although the codes generated using the RSol distribution are decodable with smaller overhead, the computational cost will be high for file sizes of practical interest. The codes generated using the distribution of equation (1), have constant encoding and decoding costs, but it is clear that an inner code is required to overcome the problem that some of the message symbols may not be covered during the encoding process. This and the study of fountain codes over mobile wireless channels is the subject of further work. Initial results indicate an additional overhead of the order of (5-10)%.

**References**:

[1] "The Global Data Transport Problem: Summary of the Digital Fountain Solution", A Digital Fountain Whitepaper, www.digitalfountain.com.

[2] Gozalvez J and Dunlop J, "Link Level Modelling Techniques for Analysing the Configuration of Link Adaptation Algorithms in Mobile Radio Networks", Proceedings of EW2004, Barcelona, February 2004.

[3] M.Luby, "LT codes", Proceedings of the 43rd Symposium on Foundation of Computer Science, 2002

[4] Amin Shokrollahi, "Raptor codes", Digital Fountain Inc., Technical Report DF2003-06-001, June 2003.

[5] David J C Mackay, "Fountain Codes", University of Cambridge.

[6] David J C Mackay, "Information Theory, Inference and Learning Algorithms", Cambridge University Press, 2003.

[7] Ravi Palanki, Jonathan S Yedidia, "Rateless codes on Noisy channels", IEEE International Symposium on Information Theory (ISIT), June 2004.

[8] Omid Etesami, Mehdi Molkaraie, Amin Shokrollahi and Digital Fountain Inc., "Raptor Codes on Symmetric Channels", Preprint 2003, http://algo.epfl.ch/index.php?p=output_pubs_XX&db=pubs/pubs_fountain.txt